\documentclass[11pt]{article}
\usepackage{bm,amsmath,color,dsfont, amssymb}
\usepackage[titletoc,toc,title]{appendix}
\usepackage{graphicx}
\usepackage[section]{placeins}
\usepackage{esint}
\usepackage{adjustbox}
\usepackage[hidelinks]{hyperref}
\usepackage{caption}
\usepackage{cleveref}
\usepackage{subcaption}
\usepackage{tabularx}
\usepackage{tgtermes} 
\usepackage{relsize}
\usepackage{booktabs}
\usepackage{wrapfig}
\usepackage{siunitx}
\usepackage{bm}
\usepackage{xcolor}
\usepackage{booktabs}
\usepackage{multirow}
\usepackage{float}

\newcommand\blue[1]{\textcolor{blue}{#1}}
\usepackage{authblk}
\begin{document}

\title{``Microstructure Modes'' -- \\ Disentangling the Joint Dynamics  of Prices \& Order Flow}

\author[1, 2]{Salma Elomari-Kessab}
\author[1, 2, 3]{Guillaume Maitrier}
\author[4]{Julius Bonart}
\author[2, 4, 5]{Jean-Philippe Bouchaud}

\affil[1]{\textit{LadHyX UMR CNRS 7646, École polytechnique, 91128 Palaiseau, France}}
\affil[2]{\textit{Chair of Econophysics and Complex Systems, École polytechnique, 91128 Palaiseau, France}}
\affil[3]{
\textit{BNP Paribas Global Markets, 20 Boulevard des Italiens, 75009 Paris, France}}
\affil[4]{\textit{Capital Fund Management, 23-25 Rue de l'Université, 75007 Paris, France.}
}
\affil[5]{\textit{Academie des Sciences, Paris 75006, France}}

\maketitle

\begin{abstract}
Understanding the micro-dynamics of asset prices in modern electronic order books is crucial for investors and regulators. In this paper, we use an order by order Eurostoxx database spanning over 3 years to analyze the joint dynamics of prices and order flow. In order to alleviate various problems caused by high-frequency noise, we propose a double coarse-graining procedure that allows us to extract meaningful information at the minute time scale. We use Principal Component Analysis to construct ``microstructure modes'' that describe the most common flow/return patterns and allow one to separate them into bid-ask symmetric and bid-ask anti-symmetric. We define and calibrate a Vector Auto-Regressive (VAR) model that encodes the dynamical evolution of these modes. The parameters of the VAR model are found to be extremely stable in time, and lead to relatively high $R^2$ prediction scores, especially for symmetric liquidity modes. The VAR model becomes marginally unstable as more lags are included, reflecting the long-memory nature of flows and giving some further credence to the possibility of ``endogenous liquidity crises''. Although very satisfactory on several counts, we show that our VAR framework does not account for the well known square-root law of price impact.  
\end{abstract}

\section{Introduction}

The micro-dynamics of asset prices is intricate, resulting from a subtle interplay between market orders, limit orders and cancellations happening at an amazingly fast pace in modern electronic order books. 
The mathematical description of the succession of these different events, the volume in the order book, and the occasional price changes when the queue at the best bid or ask empties out, is extremely difficult. This is due both to the high dimensionality of the problem, and to the presence of long-range correlations in the sign of the market/limit orders, which makes it necessary to have strong enough feedback loops. For instance, Zero Intelligence models \cite{farmer2005predictive, Bouchaud_Bonart_Donier_Gould_2018}, where agents make decisions without any strategic reasoning or foresight, fail for exactly this reason at creating coherent sequences in time.

One possible avenue, which has led to interesting results recently, is to train generative neural networks on large datasets \cite{coletta2022learning, nagy2023generative}, interpreting each event as a word and trying to guess the series of event following a given word history. Learning the underlying statistical structure of the order book dynamics would allow one to generate realistic synthetic limit order books. This would in turn offer valuable opportunities to enhance market making strategies, or its dual problem: optimal execution. It would also allow one to simulate the counter-factual impact of additional orders, that are not present in the public tape, by understanding how the market digests such orders \cite{bouchaud2009markets}. Indeed, inferring the impact of -- say -- market orders based only on the public tape is marred with conditioning problems. 

Although some success of using the analogue of Large Language Models was reported \cite{hultin2023generative, nagy2023generative, coletta2021towards}, the prediction horizon for the order book dynamics appears to be limited to a few tens of events. But since such events happen at extremely high frequencies, the time horizon of these predictions is shorter that one second for electronic liquid markets, during which the price itself seldom changes. Although possibly useful for High Frequency Trading \cite{coletta2023conditional, nagy2023generative}, one would like to develop tools that account for the joint dynamics of prices and order flows on somewhat longer time scales, say minutes. 

One of the main problems faced by ``complete'' models where all events are taken into account is that the high frequency dynamics of order books contains what one would like to call ``jitter'', i.e. orders that are placed and immediately cancelled, providing little information on the longer term fate of the order book. Another source of ``jitter'' are market orders that empty a queue at the best only to be immediately refilled by limit orders, creating high-frequency mid-point bounces. 

Our main idea in this paper is to coarse-grain and simplify the dynamics in such a way that only ``significant'' price changes (more precisely defined below) are retained. The flow of market orders, limit orders and cancellations, both at the bid and at the ask, are aggregated between two price changes and used as the relevant dynamical variables we want to focus on and predict, together with the time elapsed between two price changes and the corresponding return itself. These variables define an 8-dimensional space on which we project, in a sense, the full joint dynamics of prices and order flow. 

We then perform a Principal Component Analysis of the fluctuations, which defines ``liquidity modes'' that turn out to be stable in time and have a clean interpretation of market dynamics. This allows us to define a VAR model for predicting such modes one lag ahead, with a very significant $R^2$ score.   

One of our key findings is that one should actually distinguish between two natural coarse-graining procedures. The first one is to exclude price changes that are immediately reverted, and define other price changes as significant. However, we still see very strong mean-reversion (or ``bounce'') effects for the resulting price changes, that we call ``raw'' henceforth. We therefore define a second coarse-graining scale by aggregating $N$ successive raw price changes, constructing what we will call ``binned'' returns, choosing $N$ in such a way that the autocorrelation of successive binned returns is below $0.01$. On longer time scales, the series of price returns is thus closer to white noise, such that mechanical microstructure effects are smoothed out. For such binned data, our VAR model predicts flows with a substantial $R^2$ score ($\sim 25\%$) whereas, not unexpectedly, the prediction for returns is smaller but still significant, both in-sample and out-of-sample.

Both the ``raw'' scale and the ``binned'' scale are important for applications, but for different end users. The raw scale is presumably most useful for market makers and HFT, whereas the binned time scale is relevant for optimal execution and even, possibly, fast alpha signals. Our reduced model allows us to generate realistic time series of price changes and order flow. It also allows us to detect regime changes, when residuals with respect to the VAR model become anomalously high. 

Interestingly, when our VAR model is extended to multi-lags, we detect clear signs related to known long memory effects, i.e. several activity directions correspond to eigenvalues tending to one and become marginally stable under the dynamics. A similar effect is known to occur when one fits linear Hawkes processes to financial data \cite{bacry2015hawkes}: the only way to capture long memory is to bring the model close to instability \cite{hardiman2013critical, hardiman2014branching}. If taken at face value, the marginally stable eigenvectors of our VAR model would suggest incipient liquidity crises, a scenario advocated in various contexts, see e.g. \cite{vol_news_jp, bouchaud2011endogenous, fosset2020endogenous, marcaccioli2022exogenous, aubrun2024riding}. An alternative interpretation of such marginal stability is the effect of changing activity levels across different periods, which is in a sense another manifestation of the long range correlations of the flows. 

Finally, we can also use our model to simulate the impact of additional flows and see how far we can recover the various stylized facts reported in the literature, i.e. impact concavity and relaxation of impact after the trade is completed.  

The outline of the paper is as follows. Section \ref{Data_pres} introduces the variable of interest in our modeling. It describes the dataset, and the chosen pre-processing of the data. Section \ref{section_3} suggests an analysis of microstructure modes based on Principal Component Analysis (PCA) of our data. In Section \ref{var_model}, we present the VAR model applied to our data, with an analysis of the stability of the resulting dynamics. Measures of the price impact under our model can be found in Section \ref{section_impact}, and we conclude in Section \ref{conclusion}.

\section{Data Presentation} \label{Data_pres}

The dataset used in this study consists of 545 days of the futures contract on EuroStoxx from September 2016 to August 2019. The original data was obtained at the tick level, capturing detailed information about each price change.  During the analysis period, just under 4 million price changes were observed, with on average 7264 price changes per day. 

It is noteworthy that EuroStoxx is a liquid large-tick asset, with a spread almost always equal to one tick.  This feature puts strong constraints on possible price changes: very often the mid-point mechanically bounces back because one of the best queues is immediately refilled after its depletion. Considering that these price changes represent microstructure noise, we filtered them out of the data and retained only ``significant'' price changes, defined as follows: 

{\it A significant price change corresponds to cases when the new bid corresponds to the old ask, or when the new ask corresponds to the old bid.}

In other words, most spread-opening events correspond to a mid-point change of half a tick. If the following spread-closing event corresponds to another half-tick move in the {\it same} direction, we consider the price change to be significant. This definition allows us to remove some of the ``jitter'' that we deem insignificant in the dynamics that we want to capture, and to reduce the dimensionality of the problem by focusing on the total flux of orders between two successive price changes. 

\subsection{Variables of interest and intraday profile}

For the \(n^{th}\) significant price change of the day, occurring at time \(t_n\), we define the following variables:
\begin{align*}
    & \Delta t_n = t_n - t_{n-1} : \text{Time duration between the } (n-1)^{th} \text{ and } n^{th} \text{ price changes} \\
    & V_n^{\text{ex, a}}, V_n^{\text{ex, b}} : \text{Volume executed at the ask and bid, respectively, between } t_{n-1} \text{ and } t_n \\
    & V_n^{\text{lo, a}}, V_n^{\text{lo, b}} : \text{Volume posted to the first levels of the LOB at the bid and the ask, respectively}  \\
    & V_n^{\text{c, a}}, V_n^{\text{c, b}} : \text{Volume cancelled at the first levels of the LOB at the bid and the ask, respectively}\\
    & r_n : \text{The return generated by the price change }
\end{align*}
All variables except returns are, by definition, positive. Returns can take positive and negative values, and are equal to $\pm 1$ tick in most cases.
For later use, we stack these 8 variables into the following  8-dimensional dynamical vector 
\begin{equation} \label{eq:Xdef}
    \mathbf{X}_n = \Big(\Delta t_n, V_n^{\text{lo, b}}, V_n^{\text{lo, a}}, V_n^{\text{c, b}}, V_n^{\text{c, a}}, V_n^{\text{ex, b}}, V_n^{\text{ex, a}}, r_n\Big)
\end{equation} 
A consequence of focusing on "significant" price changes is that when the prices move up  (twice, to be deemed significant), the first inserted volume at the new, higher bid is counted in $V_n^{\text{lo, b}}$ whereas the pre-existing volume at the new ask is not - and equivalently, when the prices moves down. In other words: queues that move from a second-best to a best position are not considered as new placement flows.

\begin{figure}[htbp]
    \centering
    \includegraphics[scale=0.5]{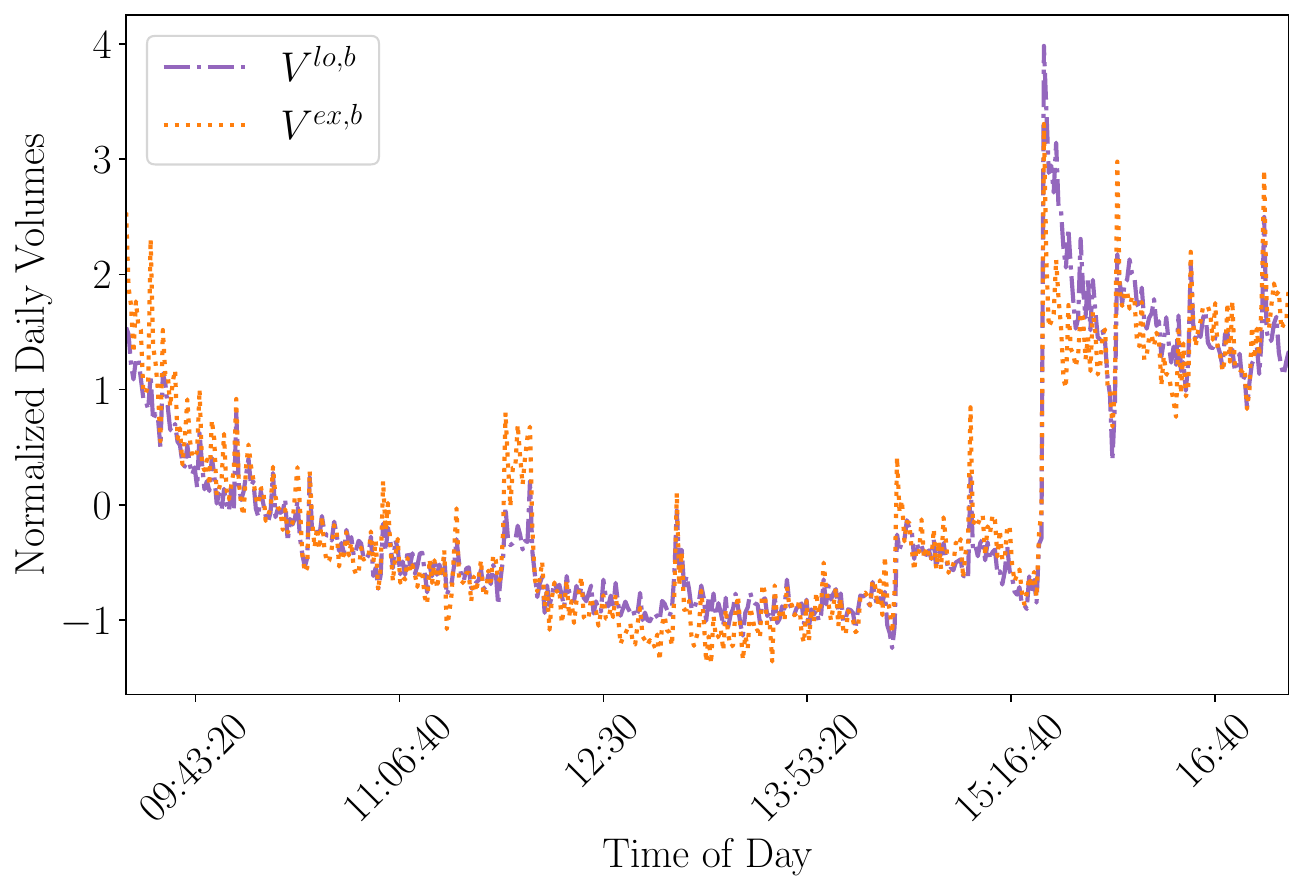}
    \caption{Normalized intraday profile of LOB placement and trade flows from the futures on EuroStoxx data. The cancellation flows have similar profiles as the placement flows and are not presented in the figure for clarity. The activity level is high at the beginning of the day and decreases until a surge of activity at the open of the U.S. market. The intraday profile is the same for all activity flows and we characterize it by a unique set of parameters given in \cref{table:exp_params}.}
    \label{fig:iday_profile}
\end{figure}

Fig. \ref{fig:iday_profile} depicts the normalized average shape of the bid volume variables $V_n^{\star,\text{ b}}$ throughout the day, binned in 1-minute intervals. The observed peak around 15:00 coincides with the opening of the US market. Since our modelling approach does not incorporate intraday volume patterns, all volumes are scaled by a smoothed average profile, fitted using two distinct exponential decay functions $A \exp(-t/\tau)+B$ with 3 parameters each: amplitude (\(A\)), on the decay time constant (\(\tau\)), and baseline (\(B\)), see \cref{table:exp_params}. 

\begin{table}[htbp]
    \centering
    \caption{Exponential decay fit parameters for the average of the 6 normalized intraday flow profiles.}
    \begin{tabular}{lccc}
        \toprule
        Parameter & Before 15:30 & After 15:30 \\
        \midrule
        Amplitude (\(A\)) & $2.20 \pm 0.14$ & $1.95 \pm 0.42$ \\
        Decay Time (\(\tau\)) & $50.0 \pm 5.7$ minutes & $6.85 \pm 2.4$ minutes \\
        Baseline (\(B\)) & $1.79 \pm 0.04$ & $3.95 \pm 0.07$ \\
        \bottomrule
    \end{tabular}
    \label{table:exp_params}
\end{table}

\subsection{A second coarse-graining}

\begin{figure}[htbp]
    \centering
    \includegraphics[scale=0.5]{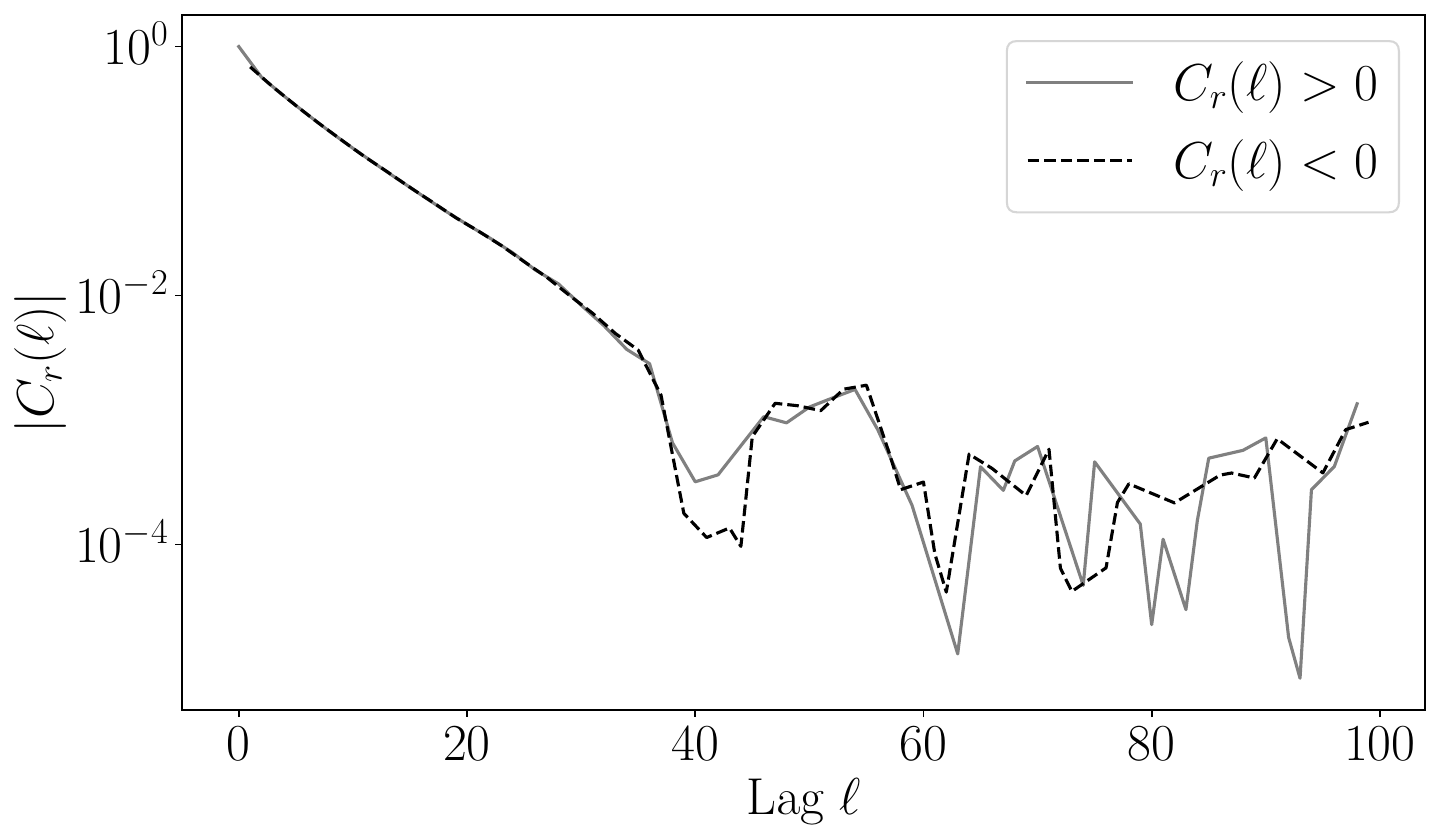}
    \caption{Absolute value of autocorrelation of returns. The alternation of positive and negative values in the autocorrelation indicates bounces in the return, which essentially disappear after binning $20$ successive price changes. }
    \label{fig:autocorrel_return}
\end{figure}

Even after filtering out price changes deemed not significant, the returns $r_n$ still show very strong anti-correlations. Fig. \ref{fig:autocorrel_return} shows that the empirical correlation function $C_r(\ell):=\langle r_n r_{n+\ell} \rangle$ can be approximated as 
\begin{equation}
    C_r(\ell) \approx (-\gamma)^\ell; \qquad (\gamma \approx 0.8 < 1).
\end{equation}
These correlations only become small $(< 0.01)$ beyond lag $\ell = 20$. To wit, strong microstructure effects still affect our ``significant'' price changes: the mid-price only becomes approximately diffusive for lags $\gtrsim 20$. 

In view of these persistent anti-correlations, we have introduced a second coarse-graining scale by further binning consecutive significant price changes into groups of 20. Throughout this paper, we will refer to our initial definition of significant price changes as ``raw'' and the aggregated (in batches of 20) price changes as ``binned''. Due to the very short time scales of the Raw Price Change data, one observes very many null flows $V_n^{\star,\text{ a}}$ or $V_n^{\star,\text{ b}}$ between $t_n$ and $t_{n+1}$, an effect that completely disappears  when the data is binned. 

\subsection{Box-Cox transformation}

Because of the strongly non-Gaussian nature of the variables $V_n$ and $\Delta t_n$, even after binning, we start by applying a ``Box-Cox'' transformation $f(x; \lambda)$ to the binned variables, with 
\begin{equation}\label{eq:boxcox}
f(x; \lambda) := \begin{cases}
\frac{{(ax)^\lambda - 1}}{{\lambda}}, & \text{if } \lambda \neq 0 \\
\log(ax), & \text{if } \lambda = 0
\end{cases}.
\end{equation}
and a parameter $\lambda$ possibly different for the volume variables $\lambda_{v}$ (for chosen to be the same for all such variables) and  $\lambda_{t}$ for the time variable. These parameters are chosen to maximise the likelihood of the Gaussian distribution of the transformed variables, which yields $\lambda _{v} = 0.20$ and  $\lambda_{t} = 0.14$.  The scale parameter $a$ can be set to unity without loss of generality.  

We will henceforth work with a series of 8-dimensional vectors $\mathbf{T}_n$ defined as:
\begin{align}
    \begin{split}
    \mathbf{T}_n = \Big( &f\left (\Delta t_n;\lambda _{\Delta t}\right), f\left(V_n^{\text{lo, b}}; \lambda _{f}\right), f\left(V_n^{\text{lo, a}}; \lambda _{f}\right), f\left(V_n^{\text{c, b}}; \lambda _{f}\right)  ,\\ &  f\left(V_n^{\text{c, a}}; \lambda _{f}\right), f\left(V_n^{\text{ex, b}}; \lambda _{f}\right), f\left(V_n^{\text{ex, a}};\lambda _{f}\right), r_n \Big).
    \end{split}
\end{align}
We further normalize the Binned Price Change data using a moving window spanning the days preceding the day of interest. Let $w$ be the width of the time window used for the computation of the means and the scales of the variables, with $w = 20$. Let $d$ be a day in the data set, and $N_k$ the number of observed price changes in day $k$. We write $\mathbf{T}_n^d$ to indicate that the vector is observed at day $d$ and define a causal local mean and the scale as follows:
\begin{equation}
    \mu_{j}^d = \frac{1}{{w}} \sum_{k=d-w}^{d-1}  \frac{1}{{N_k}}\sum_{n=1}^{N_k} (\mathbf{T}_{n}^k)_j, \quad j = 1, 2, \ldots, 8, 
\end{equation}
\begin{equation}
    (\sigma_{j}^d)^2 = \frac{1}{{w}} {\sum_{k=d-w}^{d-1} \frac{1}{{N_k}} \sum_{n=1}^{N_k}\left((\mathbf{T}_{n}^k)_j - \mu_{j}^k\right)^2}, \quad j = 1, 2, \ldots, 8,
\end{equation}
with which we normalize each component of the $\mathbf{T}_{n}$ vectors as:
\begin{equation}
    {T}_{n}^{ \prime d} = \frac{{T}^d_{n}-\mu^d}{\sigma^d}.
\end{equation}

\section{Microstructure Modes} \label{section_3}

As expected intuitively, the volume $V_n$ and time $\Delta t_n$ variables are strongly correlated. For example, a large flux of market orders might trigger more limit orders and vice-versa. It is thus natural to use a Principal Component Analysis (PCA) to understand the structure of these (same bin) correlations, and define a set of uncorrelated principal components. These vectors, ordered by their associated eigenvalues, represent the dominant microstructure modes of the market. It turns out that all these modes exhibit near-perfect bid-ask symmetry (or anti-symmetry), especially when computed using a large number of days. Since there is no reason for this symmetry to be broken at high frequencies, we manually removed all remaining spurious bid-ask asymmetry in the results presented below. Note that the PCA analysis is always performed on the Box-Cox transformed variables $T_n'$, with the averaging window $w$ chosen to be 20 days.

\subsection{PCA Analysis I: Raw Data}

\begin{figure}[htbp]
    \centering
\includegraphics[width=\linewidth]{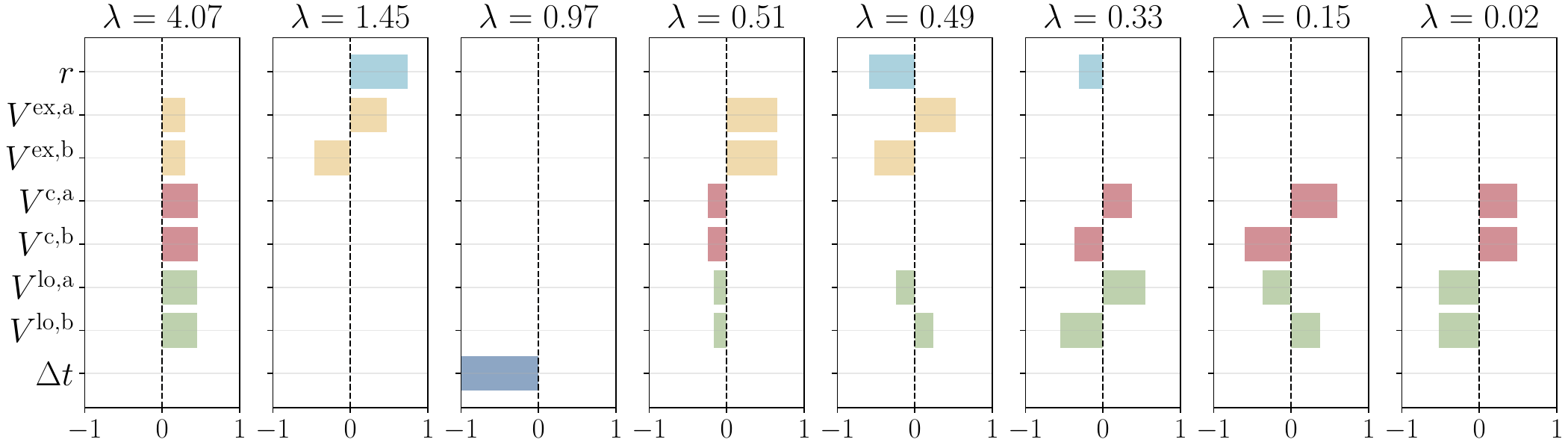}
    \caption{Normalized eigenvectors $\mathbf{U}_\alpha$ of the PCA decomposition (Raw data). For clarity purposes, the amplitudes lower than $0.15$ (corresponding to weights less than $0.15^2 \approx 2 \%$) have been set to zero. The directions should be interpreted as Box-Cox transforms of the original directions (except return $r$). }
    \label{fig:eigvectors_raw}
\end{figure}

The eigenvectors decomposition of the raw data is given in Fig. \ref{fig:eigvectors_raw}, the corresponding eigenvalues $\lambda_\alpha$ ranging from $\lambda_1=4.07$ to $\lambda_8=0.02$, with $\sum_{\alpha=1}^8 \lambda_\alpha = 8$ from the normalisation of the covariance.  

Each eigenmode $\mathbf{U}_\alpha$ has a rather intuitive and transparent interpretation, on which we comment below. Three of them are bid/ask symmetric, four are bid/ask anti-symmetric and the last one only contains duration, which appears to be independent variable at such high frequencies. Note that the {\it sign} of these eigenvectors is arbitrary; each direction is equally explored by the dynamics, with an intensity given by the square root of the corresponding eigenvalue. 

\begin{itemize}
    \item Mode 1 only contains volumes, with all coefficients positive. This represents an increase (or decrease) of general activity in the order book, with more (or less) market orders, limit orders and cancellations. It represents 51 \% of the total variance.
    \item Mode 2 mixes market order imbalance with the contemporaneous return. As expected, more executions at the ask lead to a positive return and vice versa. 
    \item Mode 3 is a pure duration mode.
    \item Mode 4 is bid/ask symmetric and describes situations where the aggressive flow becomes more active, whereas the passive flows (limit orders, cancellations) slows down -- or vice-versa.
    \item Mode 5 is anti-symmetric: more market orders at the ask than at the bid (and slightly less limit orders at the ask than at the bid), but resulting to a negative return, opposite to Mode 2. This counter-intuitive result is in this case due to the initial imbalance in the size of the queues. With the sign convention here, the bid side is less populated than the ask side, indicating net sell pressure overall. Still, higher liquidity at the ask attracts more buy market orders, explaining the excess of market orders at the ask.
    \item Mode 6, 7 and 8 are liquidity modes, since market order activity is absent from these directions. These modes represent 6.25 \% of the total variance. Mode 6 and 7 and bid/ask anti-symmetric, and Mode 8 is symmetric. Mode 7 corresponds to a growing imbalance of the available liquidity at the bid and at the ask, since we see more limit orders and less cancellations at the ask and less limit orders and more cancellations at the bid (or vice-versa). Mode 8 has a very small intensity, and corresponds to a simultaneous loss (or increase) of liquidity on both sides of the book.    
\end{itemize}

\subsection{PCA Analysis II: Binned Data} \label{eigenvectors}

We now conduct exactly the same PCA analysis but now for binned data, aggregating volume flows and returns across 20 successive price changes. The emerging eigenmodes have very much the same structure as for the raw data: the PCA yield two categories of modes, one capturing symmetric activity between the bid and ask, and the other anti-symmetric activity and non-zero price changes. 

Mode 1 again correspond to a global rise (or decline) of activity and mode 2 to a market order imbalance leading to a price change in the same direction as the imbalance. The total weight of these two modes $\lambda_1 + \lambda_2$ now reaches $\approx 6.80$, i.e. 85 \% of the total variance, compared to 69 \% for the raw data. Modes 3 and 4 are essentially the same as for raw data, apart from a permutation of their rank. The exact same thing happens for modes 5 and 6, and again for modes 7 and 8.
Mode 4 (ex mode 3 for raw data) now associates shorter time duration with more volume added and cancelled in the order book. Interestingly, bid-ask symmetric fluctuations capture 82 \% of the total variance, leaving only 18 \% of the variance to asymmetric, price changing fluctuations.

\begin{figure}[htbp]
    \centering
    \includegraphics[width=\linewidth]{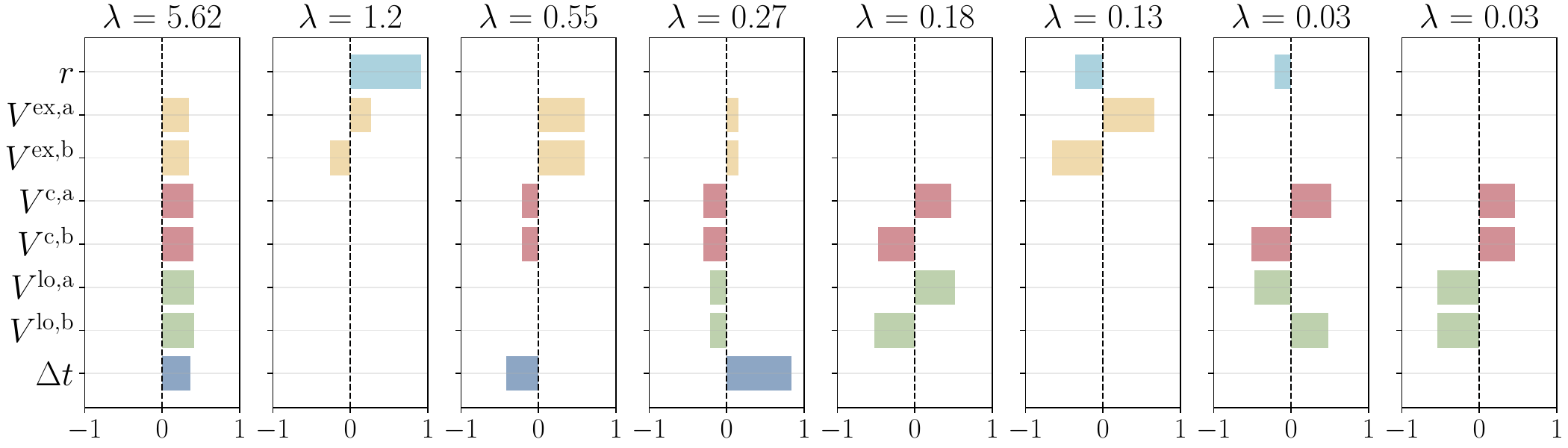}
    \caption{Normalized eigenvectors of the PCA decomposition (Binned data). For clarity purposes, the amplitudes lower than $0.15$ (corresponding to weights less than $0.15^2 \approx 2 \%$) have been set to zero. We observe 4 symmetric modes (1, 3, 4 and 8) and 4 anti-symmetric modes (2, 5, 6 and 7). The directions should be interpreted as Box-Cox transforms of the original directions (except return $r$).}
    \label{fig:eigvectors}
\end{figure}

\section{A VAR Model for Flow Dynamics} \label{var_model}

In this section, we present the mathematical framework underlying our modeling approach. We adopt a Vector Autoregression (VAR) model to capture the dynamic relationships among the variables associated with each price change. Regressing on  the projection of the data onto eigenvectors rather than directly on the data itself helps handling collinearity issues and eases the interpretability of the results. \\

We will be interested in understanding the evolution of $\mathbf{X}_n$ defined in Eq. \eqref{eq:Xdef}, for the binned data. (For the raw data, the large fraction of zero entries would require a specific treatment, following for example \cite{outcomeswithzeros, modelingzeromodifiedcount, liu2019statistical}. We leave this for later investigations). In order to do so, we transform the data using Box-Cox and set up an Auto-Regressive Vector Model in the space of the 8 principal components (or eigenmodes) described in the previous section. For every $n$, the Box-Coxed vector $\mathbf{X}_n$ is projected onto the $j^{th}$ eigenmode $\mathbf{U}_\alpha$, and the resulting projection is further demeaned and normalized to have unit variance, finally defining an 8-vector in the eigenmode space $\mathbf{Y}_n$. 

The p-lag VAR model is then specified by the following evolution equation
\begin{equation} \label{multilag-model}
\mathbf{Y}_{n} = \boldsymbol{\Phi}_1 \mathbf{Y}_{n-1} + \boldsymbol{\Phi}_2 \mathbf{Y}_{n-2} + \ldots +  \boldsymbol{\Phi}_p \mathbf{Y}_{n-p} + \mathbf{\epsilon}_n,
\end{equation}
where $\mathbf{\epsilon}_n$ represents a vector of white noise innovations and $\boldsymbol{\Phi}_k$ are $8 \times 8$ transition matrices capturing the inter-dependencies and temporal dynamics in the eigenmode space. The VAR model is calibrated using standard regression methods, except that we add by hand an additional constraint that the model has to respect the bid-ask symmetry. This means that all coefficients $(\boldsymbol{\Phi}_k)_{\alpha \beta}$ relating bid-ask symmetric modes ($\alpha = 1,3,4,8$) to bid-ask anti-symmetric modes ($\beta = 2,5,6,7$) must be zero. Without this constraint, all symmetry-breaking coefficients are found to be very small anyway.  

\subsection{1-lag VAR model}

We first focus on the $p=1$ lag VAR model:
\begin{equation} \label{1_lag_model}
\mathbf{Y}_{n} = \boldsymbol{\Phi}_1 \mathbf{Y}_{n-1} + \mathbf{\epsilon}_n.
\end{equation}

\begin{figure}[h!] 
    \centering
    \includegraphics[width=\linewidth]{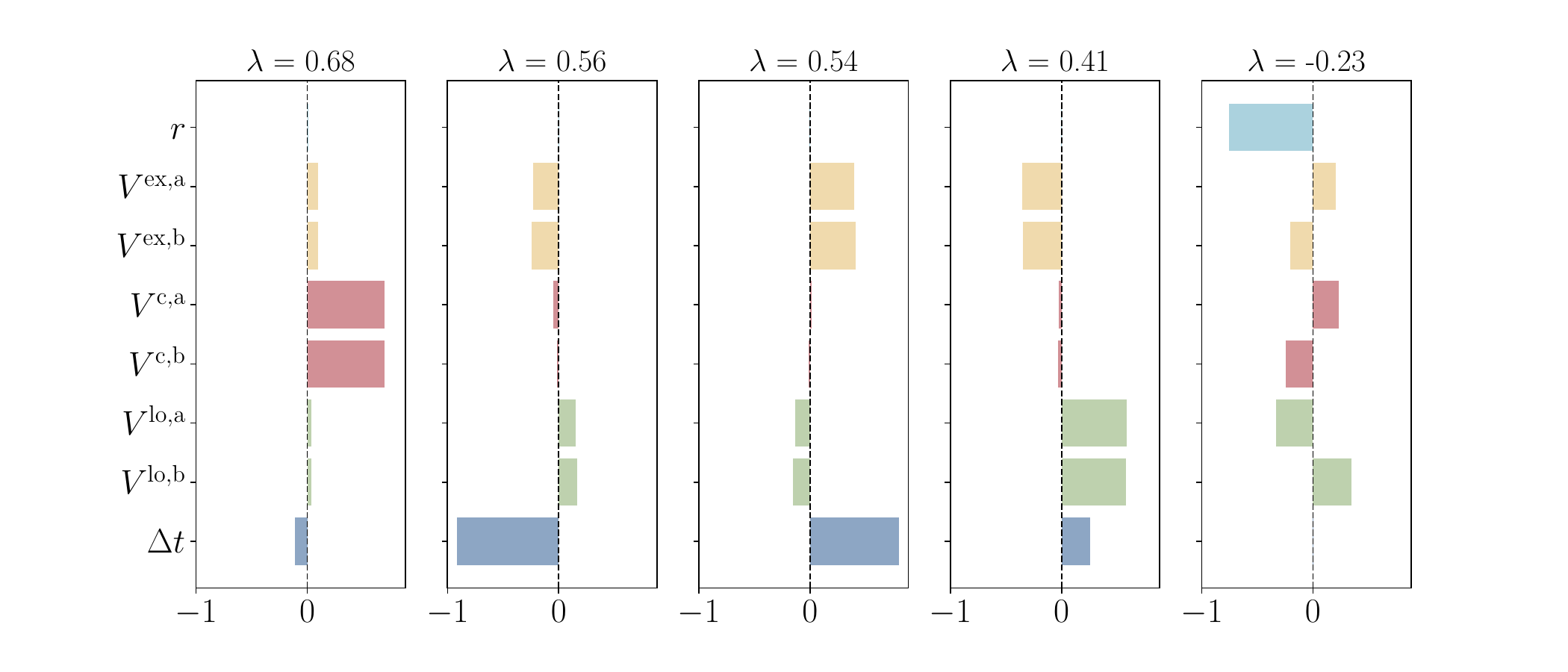}
    \caption{5 eigenvectors with the largest eigenvalue norm from the decomposition of the transition matrix. The 4 first eigenvectors have positive eigenvalues and describe symmetric scenarios in the bid and the ask, the fifth one is anti-symmetric and mean-reverting. }
    \label{fig:5_eigenvectors}
\end{figure}

The transition matrix $\boldsymbol{\Phi}_1$ is presented in the \cref{tab:transition_matrix_mode}. The most significant elements, i.e., such that $|(\boldsymbol{\Phi}_1)_{\alpha \beta}| > 0.1$, are highlighted in bold and correspond mostly to diagonal elements (except $22$ and $66$). However, a better description of the transition matrix is in terms of its eigenvalues and eigenvectors. 6 eigenvectors correspond to real eigenvalues, 5 positive and one negative, and 2 eigenvectors correspond to a pair of complex conjugate eigenvalues, with a very small modulus. The five eigenvectors with largest norm are shown in Fig. \ref{fig:5_eigenvectors}. The fact that all eigenvalues within the unit circle means that the lag-1 VAR model is stable, with fluctuations dampening instead of getting amplified.  Notice that the top eigenvalue is equal to $0.68$ and corresponds to a symmetric cancellation mode, mostly reflecting the activity of market makers. 

The second mode, with eigenvalue $0.56$, is also symmetric and corresponds to more limit orders, less market orders and less inter price change time, or vice-versa. The largest anti-symmetric mode has eigenvalue $\lambda_5=-0.23$ and is the imbalance level for all flows, which is seen to be mean-reverting (since $\lambda_5 < 0$).


\begin{table}[h!]
\centering
\begin{tabular}{l|cccccccc}
\toprule
\textbf{Mode} & \textbf{1S} & \textbf{2A} & \textbf{3S} & \textbf{4S} & \textbf{5A} & \textbf{6A} & \textbf{7A} & \textbf{8S} \\
\midrule
\textbf{1S} & \textbf{0.56} & 0.00 & -0.02 & -0.03 & 0.00 & 0.00 & 0.00 & 0.09 \\
\textbf{2A} & 0.00 & -0.06 & 0.00 & 0.00 & -0.00 & 0.04 & -0.09 & 0.00 \\
\textbf{3S} & 0.05 & 0.00 & \textbf{0.53} & 0.00 & 0.00 & 0.00 & 0.00 & -0.09 \\
\textbf{4S} & -0.06 & 0.00 & -0.02 & \textbf{0.59} & 0.00 & 0.00 & 0.00 & -0.03 \\
\textbf{5A} & 0.00 & 0.01 & 0.00 & 0.00 & \textbf{0.15} & -0.02 & -0.04 & 0.00 \\
\textbf{6A} & 0.00 & 0.08 & 0.00 & 0.00 & -0.09 & -0.05 & 0.09 & 0.00 \\
\textbf{7A} & 0.00 & -0.08 & 0.00 & 0.00 & -0.08 & 0.09 & \textbf{-0.11} & 0.00 \\
\textbf{8S} & \textbf{0.12} & 0.00 & -0.05 & -0.04 & 0.00 & 0.00 & 0.00 & \textbf{0.52} \\
\bottomrule
\end{tabular}

\caption{The transition matrix for microstructure modes, where values exceeding a significance threshold of $0.05$ in the corresponding p-value have been set to zero. Columns correspond to input modes from time $n-1$, rows to predicted modes at time $n$. Symbol $S$ (or $A$) refers to the bid-ask symmetry (anti-symmetry) of the modes. }
\label{tab:transition_matrix_mode}
\end{table}

The success of the lag-1 VAR model can be quantified in terms of the predictive $R^2$ scores, presented in \cref{tab:r2_scores}, both across modes of the transformed variables $\mathbf{Y}_n$ and for the original variables $\mathbf{X}_n$. Note that, as expected, $R^2$ scores are much higher ($\sim 0.28 - 0.32$) for symmetric modes, which carry no information on returns, than for anti-symmetric modes ($\sim 0.01 - 0.03$). However, the in-sample $R^2$ scores and out-of-sample scores are close, highlighting the fact that the predictive value of the VAR model is statistically significant. We have used the first 465 days of the data for model calibration and computing the in-sample scores. The remaining 80 days are allocated for computing the out-of-sample scores.  

\begin{table}[h!]
\centering
\begin{tabular}{l|cccccccc}
\toprule
\textbf{Mode} & \textbf{1S} & \textbf{2A} & \textbf{3S} & \textbf{4S} & \textbf{5A} & \textbf{6A} & \textbf{7A} & \textbf{8S} \\
\midrule
\textbf{In Sample (\%)} & 32.4 & 1.2 & 29.1 & 35.1 & 2.43 & 2.39 & 3.07 & 28.8 \\
\midrule
\textbf{Out Of Sample (\%)} & 28.0 & 1.11 & 21.8 & 36.1 & 4.33 & 1.68 & 2.24 & 32.5 \\
\bottomrule
\toprule
\textbf{Variable} & \textbf{$\Delta t$} & \textbf{$V^{\text{lo, b}}$} & \textbf{$V^{\text{lo, a}}$} & \textbf{$V^{\text{c, b}}$} & \textbf{$V^{\text{c, a}}$} & \textbf{$V^{\text{ex, b}}$} & \textbf{$V^{\text{ex, a}}$} & \textbf{$r$} \\
\midrule
\textbf{In Sample (\%)} & 21.3 & 29.8 & 29.8 & 36.4 & 36.0 & 25.3 & 24.8 & 1.60 \\
\midrule
\textbf{Out Of Sample (\%)} & 27.4 & 22.7 & 21.5 & 25.6 & 24.1 & 22.9 & 25.0 & 1.46 \\
\bottomrule
\end{tabular}
\caption{$R^2$ scores in $\%$ both in mode space (top) and in the original space (bottom). Symbol $S$ (or $A$) refer to the bid-ask symmetry (anti-symmetry) of the modes. }
\label{tab:r2_scores}
\end{table}

The $R^2$ score of $1.6 \%$ for return $r$ is of particular interest. It is in particular significantly higher than the score of $0.49 \%$ obtained when predicting returns using the past return as the only feature. This shows that flow variables add useful predictive power to the return variable.


\subsection{Multi-lag VAR model}

In this subsection, we extend our modeling approach to the multi-lag Vector Autoregression VAR(p) model, specified by Eq. \eqref{multilag-model} with $p > 1$. 
Interestingly, adding more lags reduces auto- correlation of residuals and increases the out of sample $R^2$ score of all the modes, by $\sim 25 \%$ both for the symmetric and anti-symmetric ones when $p$ increases from 1 to 10 -- see \cref{tab:scores,tab:r2_scores_8_lags}.


\begin{table}[h!]
\begin{adjustbox}{width=\columnwidth,center}
\begin{tabular}{l|cccccccccc}
\toprule
\textbf{Lags} & \textbf{1} & \textbf{2} & \textbf{3} & \textbf{4} & \textbf{5} & \textbf{6} & \textbf{7} & \textbf{8} & \textbf{9} & \textbf{10} \\
\midrule
\textbf{In Sample S (\%)} & 31.4 & 35.8 & 37.3 & 38.1 & 38.5 & 38.8 & 39.0 & 39.2 & 39.3 & 39.4 \\
\textbf{Out Of Sample S (\%)} & 29.6 & 34.3 & 36.2 & 37.0 & 37.3 & 37.6 & 37.9 & 38.1 & 38.3 & 38.4 \\
\textbf{In Sample A (\%)} & 2.29 & 2.56 & 2.69 & 2.77 & 2.86 & 2.94 & 3.00 & 3.05 & 3.10 & 3.15 \\
\textbf{Out Of Sample A (\%)} & 2.35 & 2.56 & 2.65 & 2.73 & 2.79 & 2.85 & 2.92 & 2.99 & 3.03 & 3.04 \\
\bottomrule
\end{tabular}
\end{adjustbox}
\caption{Average $R^2$ scores for symmetric (S) and asymmetric (A) modes in-sample and out-of-sample for different number of lags $p$.}
\label{tab:scores}
\end{table}

\begin{table}[h!]
    \centering
    \begin{tabular}{l|cccccccc}
        \toprule
        \textbf{Mode} & \textbf{1S} & \textbf{2A} & \textbf{3S} & \textbf{4S} & \textbf{5A} & \textbf{6A} & \textbf{7A} & \textbf{8S} \\
        \midrule
        \textbf{In Sample (\%)} & 36.0 & 1.46 & 35.5 & 43.5 & 3.47 & 2.29 & 4.00 & 36.3 \\
        \midrule
        \textbf{Out Of Sample (\%)} & 36.7 & 1.30 & 27.5 & 45.4 & 5.53 & 2.22 & 2.89 & 42.6\\
        \bottomrule
        \toprule
        \textbf{Variable} & \textbf{$\Delta t$} & \textbf{$V^{\text{lo, b}}$} & \textbf{$V^{\text{lo, a}}$} & \textbf{$V^{\text{c, b}}$} & \textbf{$V^{\text{c, a}}$} & \textbf{$V^{\text{ex, b}}$} & \textbf{$V^{\text{ex, a}}$} & \textbf{$r$} \\
        \midrule
        \textbf{In Sample (\%)} & 30.4 & 37.6 & 37.9 & 44.0 & 43.8 & 32.1 & 31.5 & 1.87 \\
        \midrule
        \textbf{Out Of Sample (\%)} & 36.3 & 31.4 & 29.4 & 34.1 & 32.1 & 28.4 & 31.3 & 1.7 \\
        \bottomrule
    \end{tabular}
    \caption{$R^2$ scores using the VAR(8) in $\%$ both in mode space (top) and in the original space (bottom). Note that the out-of-sample $R^2$ score of the returns increases from $1.46$ for $p=1$ to $1.7$ for $p=8$.  }
    \label{tab:r2_scores_8_lags}
\end{table}
Another interesting question is whether adding memory to the system makes it less stable. In order to discuss this point, let us look for a vector $\mathbf{Z}$ such that at long times the p-VAR model in the absence of innovations would yield 
\[ \mathbf{Y}_n \approx_{n \gg 1} \gamma^n \mathbf{Z}\].

Injecting in \cref{multilag-model} and dividing by $\gamma^n$, we find the following condition:
\begin{equation}
    \mathbf{Z} = \mathbb{M}_p \mathbf{Z}, \qquad \mathbb{M}_p(\gamma):= \left[\sum_{k=1}^p \gamma^{-k} \mathbf{\Phi}_k \right].
\end{equation}
In other words, one should look for a value of {$\gamma$ } such that the matrix $\mathbb{M}_p (\gamma)$ has one eigenvalue exactly equal to unity, the corresponding eigenvector defining $\mathbf{Z}$. The least stable direction of the p-VAR model is associated with the largest possible value of $|\gamma|$ (with $\gamma$ possibly complex). 

\begin{figure}[h!]
    \centering
    \includegraphics[width = 0.75\textwidth]{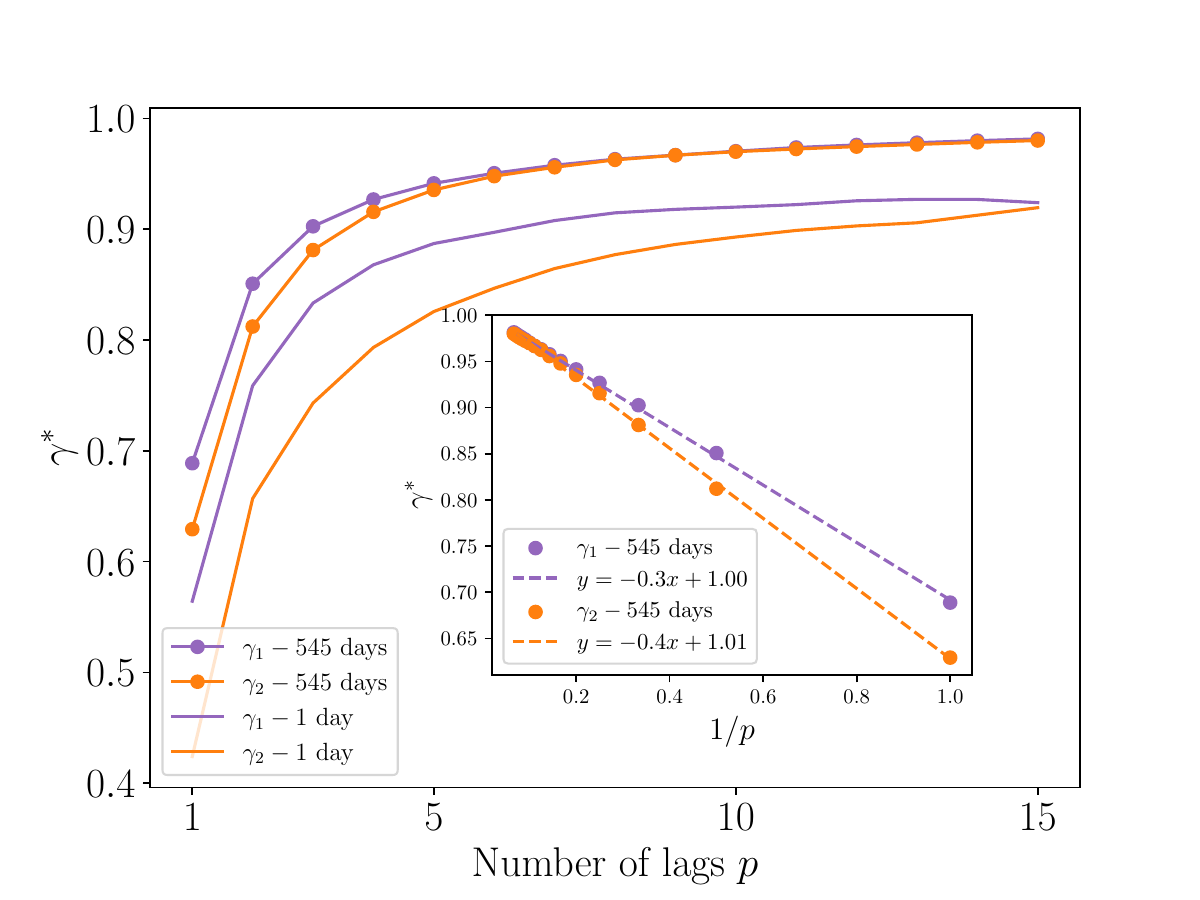}
    \caption{The two largest values of $\gamma_{1,2}(p)$, such as the matrix $\mathbb{M}_p(\gamma)$ has an eigenvalue equal to unity, as a function of lag $p$. The plot compares methods for normalizing the data: one using all 545 available days and the other where each day is normalized independently. Inset: same results, plotted as a function of $1/p$ showing a near perfect linear behaviour extrapolating to unity when $p \to \infty$. }
    \label{fig:beta}
\end{figure}

\begin{figure}[h!]
    \centering
    \includegraphics[width = \linewidth]{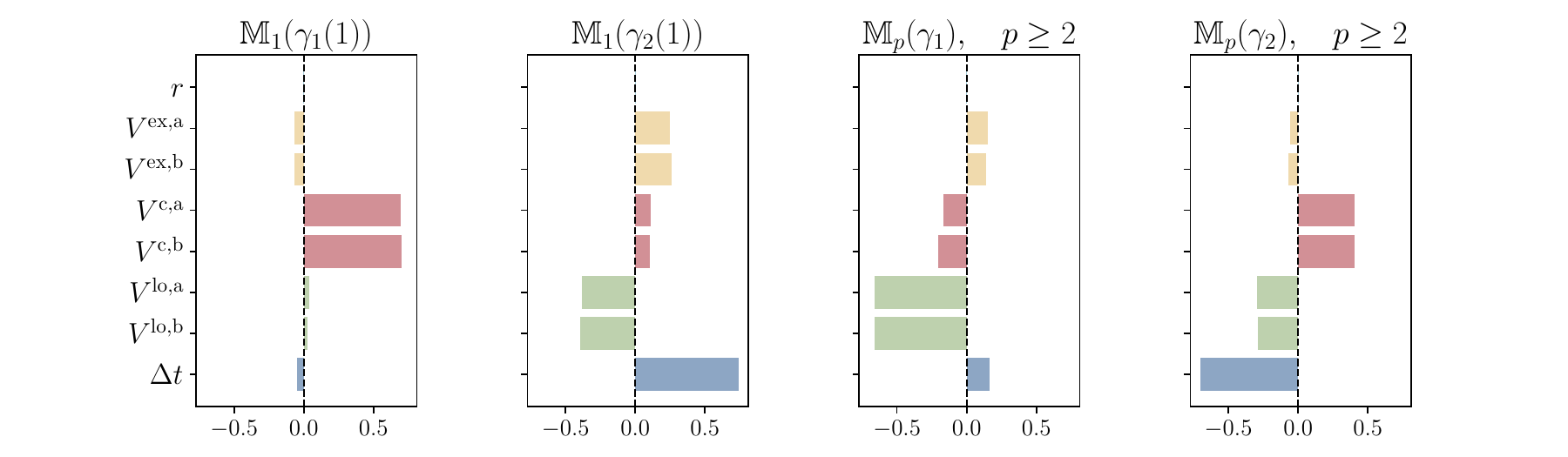}
    \caption{Dominant eigenvectors of $\mathbb{M}_p(\gamma_{1,2})$. For $p = 1$ we recover eigenvectors from Fig. \ref{fig:5_eigenvectors}. When $p \geq 2$, all dominant eigenvectors are essentially independent of $p$ and are associated with important liquidity fluctuations. For example $\gamma_2$ describes a persistent mode with less order placements and more cancellations, which can lead to liquidity crises.}
    \label{fig:beta 0 rank 0}
\end{figure}

Quite interestingly, we observe in Fig. \ref{fig:beta} that both $\gamma_1(p)$ and $\gamma_2(p)$ can be fitted as $1 - C_{1,2}/p$ and therefore appear converge to unity as the number of lags increases. This means that the dominant eigenvectors, shown in Fig. \ref{fig:beta 0 rank 0}, become more and more persistent as we increase the number of lags $p$. This suggests that the flow dynamics is in fact {\it marginally stable}, which is in line with the well-known stylized fact that order flow has power law, long memory correlations \cite{bouchaud2009markets}, corresponding to a unit root within a VAR description, or to marginal stability within a Hawkes process description \cite{hardiman2014branching}. Marginal stability could however result from the inadequacy of the VAR model to represent the data, since the only way to represent long memory correlations within a VAR framework is to have unit roots.  

Dominant eigenvectors are identical for all $p\geq 2$ and describe liquidity fluctuations. The mode associated with $\gamma_1(p)$ predicts less (or more) placements than usual. The second one, with rate $\gamma_2(p)$, describes a persistent mode with less order placements and more cancellations, which can lead to liquidity crises, as argued in \cite{fosset2020endogenous}. Even if $\gamma_2(p)$ is below unity, the system appears to be very close to this stability boundary, and therefore be prone to endogenous liquidity crisis. In this context, recall that we chose a particularly stable, large tick contract (the Eurostoxx); it would be interesting to perform the same analysis with small tick single name stocks.

\section{An Attempt to Model Price Impact}\label{section_impact}  

In this section, we are interested in understanding the impact of the trading of one agent on the market and its future states within the VAR framework established above. For the rest of the section, we focus on the impact of perturbations of the flows at the ask without any loss of generality since the regression matrix is symmetric between the bid and the ask. 

A phenomenon commonly studied in the literature is price impact \cite{bouchaudprice, Bouchaud_Bonart_Donier_Gould_2018, webster2023handbook}, or by how much a trader modifies the price of an asset by buying or selling it. This metric is crucial for practitioners, but also from an academic point of view. Price impact exhibits interesting theoretical properties, such as the so-called square root law (for a review, see \cite{Bouchaud_Bonart_Donier_Gould_2018, webster2023handbook}).

In principle, the mechanical impact of market orders (i.e. the part that is independent of any information motivating the trades) is defined as \cite{Bouchaud_Bonart_Donier_Gould_2018, webster2023handbook}
\begin{equation}\label{eq:reaction impact}
    \mathcal{I}(\ell) := \mathbb{E} [m_{t+\ell} - m_t | \text{exec}_t ] - \mathbb{E} [m_{t+\ell} - m_t| \text{no-exec}_t ],
\end{equation}
where $m(t)$ is the mid-price at time $t$, when a market order is executed. In other words, one should compare the price change between time $t$ and $t+\ell$ with and without order execution. Of course, such a measurement is impossible, since these two states of the market are mutually exclusive. Therefore, in practice one assumes that for short enough time scales, market orders issued by slow traders have little short term predictability such that the second term in Eq. \eqref{eq:reaction impact} is negligible. Hence the observable impact is defined as
\begin{equation}
    \mathcal{I}^{\text{obs}}(\ell|exec_t) := \mathbb{E} [m_{t+\ell} - m_t].
\end{equation}
The whole idea of constructing a faithful generating model for prices and order flow is to be able to perform numerically the ``do-operation'' \cite{webster2023handbook} described in Eq.   \eqref{eq:reaction impact}. 

We have performed such a numerical experiment using the VAR model calibrated above on binned data -- which, we recall, aggregates together 20 successive significant price changes. The procedure is as follows: we add to the observed flow of market orders at the ask a specific quantity corresponding to our extra buyer, between coarse-grained time $n$ and time $n+k$. At each time step, the instantaneous impact is calculated using the average impact curves obtained in \cite{patzelt2018universal}, that are reproduced in  the \ref{appendix_impact}. 

However, there is a subtlety related to the execution flows predicted by the VAR model. Rotating the matrix $\Phi_1$ into real flows' space, we obtain the matrix shown in \cref{tab:mat_real_space}. 
\begin{table}[h!]

\centering
\renewcommand{\arraystretch}{1.2}
\begin{tabular}{l|cccccccc}
\toprule
\textbf{Variables} & $\Delta t$ & $ V^{\text{lo, b}}$ & $V^{\text{lo, a}}$ & $V^{\text{c, b}}$ & $V^{\text{c, a}}$ & $V^{\text{ex, b}}$ & $ V^{\text{ex, a}}$ & $r$ \\
\midrule
$\Delta t$ & 0.56 & 0.05 & -0.05 & -0.00 & 0.00 & -0.02 & \textbf{-0.03} & -0.0 \\
$ V^{\text{lo, b}}$ & -0.02 & 0.21 & 0.22 & 0.06 & -0.05 & 0.02 & \textbf{-0.01} & 0.06 \\
$V^{\text{lo, a}}$ & -0.02 & 0.22 & 0.21 & -0.05 & 0.06 & -0.01 & \textbf{0.02} & -0.05 \\
$V^{\text{c, b}}$ & -0.00 & 0.08 & -0.06 & 0.39 & 0.28 & -0.00 & \textbf{0.02} & -0.04 \\
$V^{\text{c, a}}$ & -0.00 & -0.06 & -0.08 & 0.28 & 0.39 & 0.06 & \textbf{-0.00} & 0.04 \\
$V^{\text{ex, b}}$ & 0.02 & 0.02 & 0.04 & -0.02 & 0.04 & 0.26 & \textbf{0.28} & -0.05 \\
$ V^{\text{ex, a}}$ & 0.02 & 0.04 & 0.02 & 0.04 & -0.02 & 0.28 & \textbf{0.26} & 0.05 \\
$r$ & -0.00 & 0.06 & -0.05 & -0.06 & 0.04 & -0.02 & \textbf{0.02} & -0.14 \\
\bottomrule
\end{tabular}

\caption{An approximation of the transition matrix in the real flows' space obtained as a rotation of $\Phi_1$ back to the real variables space. Columns correspond to input variables at time step $n-1$, rows correspond to an estimation of the variables at time step $n$. Note: This is not strictly speaking a transition matrix because of the non-linear Box-Cox operation.}
\label{tab:mat_real_space}
\end{table}
This matrix reveals that an increase in the market order flow at the ask is most likely followed at the next time step by an increase in market order flows in both the ask and the bid, with slightly higher values observed at the opposite side. The succession of market orders at the same side is a manifestation of the well-known long range correlation of the flows in the market \cite{bouchaud2009markets}, which is primarily due to metaorder splitting, with very little contribution from herding \cite{toth2015equity}. 

\begin{figure}[htbp] 
\centering
  \centering
  \includegraphics[width=0.9\linewidth]{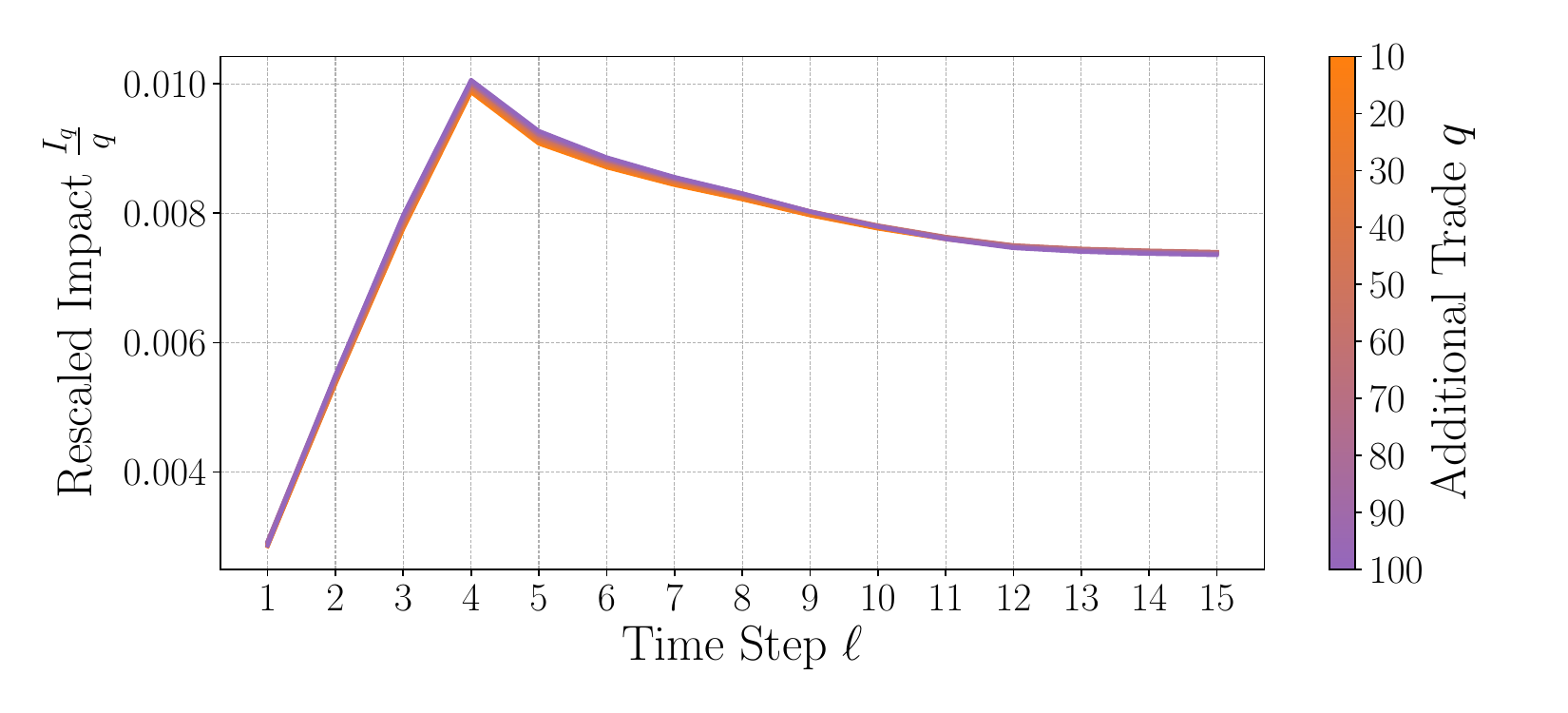}
\caption{Simulations of impact of metaorders of length $k=4$. For the first 4 time steps, a market order of size $q$ is added to the observed execution flow. Starting from the $5^{th}$ time step the market is no longer perturbed. We rescale the measured impact by the size of the added trade flow.}
\label{fig:impact_meta_order}
\end{figure}

Our model has been trained on real-world price and flow data, whose causal structure includes but cannot be reduced to a perturbation-response mechanism. Single market participants do not act in isolation and they may, through complex trading strategies, influence the market dynamics on long time scales, and even cross-sectionally. To the extent that the exogenous perturbation, whose impact we wish to simulate, is not representative of the average market participants' trading schedule, the model cannot fully distinguish whether correlations are due to market response or individual complex trading strategies. In order to model consistently the impact of a specific exogenous metaorder, we must avoid double counting such contributions. Thus, as an approximation, within our simulation framework we disregard subsequent execution orders predicted by the model on the same side and only take into account induced effects. The perturbed flows and returns are then propagated forward in time using the VAR model. The total price impact is then obtained by subtracting the unperturbed observed price trajectory and averaging over time. However, it is important to note that this approach is, even on a conceptual level, an approximation whose accuracy is difficult to quantify. The result for the VAR(10) model is shown in Fig. \ref{fig:impact_meta_order}. 

Empirically, as mentioned above, impact is strongly concave, and shows a square-root dependence both in time (within a metaorder) and in total size (at peak impact), see \cite{Bouchaud_Bonart_Donier_Gould_2018}, chapter 12. Furthermore, such an impact strongly mean-reverts at the end of a metaorder. Our methodology, however, yields impact behaviour that differs in notable ways. What we observe in Fig. \ref{fig:impact_meta_order} is that the generated impact is only slightly concave within the metaorder, and then decays back down once the metaorder is completed. Despite this, the peak impact is linear in the size of the metaorder, contrarily to the concave behavior in observed market data. This linear shape is in fact expected within our perturbation approach, where the added trade flow is small enough to be absorbed by the market, leading to a linear behavior of the non-linear Box-Cox transformation. Furthermore, the level of reversion of the price between the moment we stop perturbing the market and when the price stabilizes is around $75\%$ of the peak impact, whereas impact decay is much steeper in real data, with a significantly lower plateau value \cite{bucci2018slow}.

The conclusion of this section is that although our VAR framework offers a good benchmark for modelling the impact of metaorders, a crucial element appears to be missing since the strongly concave, mean-reverting nature impact is missed. We conjecture that such a missing element is an explicit reference to recent price changes, in a way to incorporate the idea of asymmetric latent liquidity, as argued in \cite{donier2015fully, Bouchaud_Bonart_Donier_Gould_2018}. Additionally, due to the use of the impact curves from \cite{patzelt2018universal} for computing the instantaneous return, the return produced by our model is diluted in scale. To address these limitations, future work could benefit from exploring models with enhanced non-linearity, such as neural networks.

\section{Conclusion and Further Discussions} \label{conclusion}

The frantic and noisy order book dynamics at the highest frequency hamper modelling attempts based on order by order activity. In this work, we have devised a specific coarse-graining procedure to extract meaningful information from such erratic flow data. First, in order to remove ``flickering'' bid-ask bounce noise, we have proposed a definition of significant price changes, 
and defined the flow variables of interest as aggregates of market orders, limit orders and cancellations between two such significant price changes. 

However, we have found it necessary to introduce a second coarse-graining time scale in order to (i) smooth out strong price mean-reversion that survives until $\sim 20$ significant price changes and (ii) eliminate the large quantity of zeros in the flow variables that make linear analysis difficult to interpret.

One of our most interesting novel result is the appearance of what we called ``microstructure modes'', i.e. principal components of the joint, coarse-grained dynamics of price and order flow. These modes are extremely stable over time and all have an intuitive interpretation. They fall into two categories: bid-ask symmetric and bid-ask anti-symmetric. The first category describes, for example, an increase/decrease of cancellations and a decrease/increase of limit orders on both sides of the book simultaneously, associated to the dynamics of liquidity. The second category describes, for example, an increase of market orders at the ask and a decrease of market orders at the bid, associated to a positive price return. 

Using these microstructure modes as inputs, we built and calibrated a multi-lag VAR model that captures their dynamics. The model is stable in time and leads to high $R^2$ scores $\sim 30-40 \%$ for symmetric modes and, as expected, lower but significant $R^2$ scores $\sim 2-3 \%$ for anti-symmetric (directional) modes. Non-linear, neural network models that take our 
microstructure modes as features should improve further the quality of the prediction.

We have found that the VAR model becomes marginally stable as the number of lags increases. This reflects the well-known long memory nature of the order flow in financial markets. The analysis of the flow directions that become unstable gives further credence to the ``endogenous liquidity crisis'' scenario suggested in \cite{vol_news_jp, bouchaud2011endogenous, fosset2020endogenous, marcaccioli2022exogenous, aubrun2024riding}.   

Finally, we have used our VAR formalism to measure the impact of metaorders on the price. Although we observe some price mean-reversion at the end of the metaorder, similar to real data, we failed to reproduce the concave square-root dependence of impact on time and volume. We conjectured that an explicit conditioning of the VAR transition matrix on the recent returns is needed to capture ``latent liquidity'' effects that are thought to be at the origin of impact concavity \cite{donier2015fully, Bouchaud_Bonart_Donier_Gould_2018}.  

When working on the ``raw", unbinned data, we were confronted with the fact that at short time scales, most of the observed flow volumes are null, making our linear VAR model unsuitable. One could address this problem using recent statistical techniques \cite{outcomeswithzeros, modelingzeromodifiedcount, liu2019statistical}, or using more complex neural network architectures combining recurrent neural networks and attention techniques. It would also be interesting to revisit the price impact problem within this framework.

\section*{Acknowledgments}
We would like to thank Nirbhay Patil, C\'ecilia Aubrun and J\'erome Garnier-Brun for their helpful suggestions. This research was conducted within the Econophysics \& Complex Systems Research Chair, under the aegis of the Fondation du Risque, the Fondation de l'\'Ecole Polytechnique and Capital Fund Management.
\bibliographystyle{unsrt}
\bibliography{references}  

\begin{thebibliography}{10}

\bibitem{farmer2005predictive}
J~Doyne Farmer, Paolo Patelli, and Ilija~I Zovko.
\newblock The predictive power of zero intelligence in financial markets.
\newblock {\em Proceedings of the National Academy of Sciences},
  102(6):2254--2259, 2005.

\bibitem{Bouchaud_Bonart_Donier_Gould_2018}
Jean-Philippe Bouchaud, Julius Bonart, Jonathan Donier, and Martin Gould.
\newblock {\em The Impact of Market Orders}, page 208–228.
\newblock Cambridge University Press, 2018.

\bibitem{coletta2022learning}
Andrea Coletta, Aymeric Moulin, Svitlana Vyetrenko, and Tucker Balch.
\newblock Learning to simulate realistic limit order book markets from data as
  a world agent.
\newblock In {\em Proceedings of the third acm international conference on ai
  in finance}, pages 428--436, 2022.

\bibitem{nagy2023generative}
Peer Nagy, Sascha Frey, Silvia Sapora, Kang Li, Anisoara Calinescu, Stefan
  Zohren, and Jakob Foerster.
\newblock Generative ai for end-to-end limit order book modelling: A
  token-level autoregressive generative model of message flow using a deep
  state space network.
\newblock In {\em Proceedings of the Fourth ACM International Conference on AI
  in Finance}, pages 91--99, 2023.

\bibitem{bouchaud2009markets}
Jean-Philippe Bouchaud, J~Doyne Farmer, and Fabrizio Lillo.
\newblock How markets slowly digest changes in supply and demand.
\newblock In {\em Handbook of financial markets: dynamics and evolution}, pages
  57--160. Elsevier, 2009.

\bibitem{hultin2023generative}
Hanna Hultin, Henrik Hult, Alexandre Proutiere, Samuel Samama, and Ala
  Tarighati.
\newblock A generative model of a limit order book using recurrent neural
  networks.
\newblock {\em Quantitative Finance}, 23(6):931--958, 2023.

\bibitem{coletta2021towards}
Andrea Coletta, Matteo Prata, Michele Conti, Emanuele Mercanti, Novella
  Bartolini, Aymeric Moulin, Svitlana Vyetrenko, and Tucker Balch.
\newblock Towards realistic market simulations: a generative adversarial
  networks approach.
\newblock In {\em Proceedings of the Second ACM International Conference on AI
  in Finance}, pages 1--9, 2021.

\bibitem{coletta2023conditional}
Andrea Coletta, Joseph Jerome, Rahul Savani, and Svitlana Vyetrenko.
\newblock Conditional generators for limit order book environments:
  Explainability, challenges, and robustness.
\newblock In {\em Proceedings of the Fourth ACM International Conference on AI
  in Finance}, pages 27--35, 2023.

\bibitem{bacry2015hawkes}
Emmanuel Bacry, Iacopo Mastromatteo, and Jean-Fran{\c{c}}ois Muzy.
\newblock Hawkes processes in finance.
\newblock {\em Market Microstructure and Liquidity}, 1(01):1550005, 2015.

\bibitem{hardiman2013critical}
Stephen~J Hardiman, Nicolas Bercot, and Jean-Philippe Bouchaud.
\newblock Critical reflexivity in financial markets: a hawkes process analysis.
\newblock {\em The European Physical Journal B}, 86:1--9, 2013.

\bibitem{hardiman2014branching}
Stephen~J Hardiman and Jean-Philippe Bouchaud.
\newblock Branching-ratio approximation for the self-exciting hawkes process.
\newblock {\em Physical Review E}, 90(6):062807, 2014.

\bibitem{vol_news_jp}
Armand Joulin, Augustin Lefevre, Daniel Grunberg, and Jean-Philippe Bouchaud.
\newblock Stock price jumps: news and volume play a minor role.
\newblock {\em Wilmott Magazine}, 46, 2008.

\bibitem{bouchaud2011endogenous}
Jean-Philippe Bouchaud.
\newblock The endogenous dynamics of markets: Price impact, feedback loops and
  instabilities.
\newblock {\em Lessons from the credit crisis}, pages 345--74, 2011.

\bibitem{fosset2020endogenous}
Antoine Fosset, Jean-Philippe Bouchaud, and Michael Benzaquen.
\newblock Endogenous liquidity crises.
\newblock {\em Journal of Statistical Mechanics: Theory and Experiment},
  2020(6):063401, 2020.

\bibitem{marcaccioli2022exogenous}
Riccardo Marcaccioli, Jean-Philippe Bouchaud, and Michael Benzaquen.
\newblock Exogenous and endogenous price jumps belong to different dynamical
  classes.
\newblock {\em Journal of Statistical Mechanics: Theory and Experiment},
  2022(2):023403, 2022.

\bibitem{aubrun2024riding}
Cecilia Aubrun, Rudy Morel, Michael Benzaquen, and Jean-Philippe Bouchaud.
\newblock Riding wavelets: A method to discover new classes of price jumps.
\newblock {\em arXiv preprint arXiv:2404.16467}, 2024.

\bibitem{outcomeswithzeros}
Aaron~J. Boulton and Anne Williford.
\newblock Analyzing skewed continuous outcomes with many zeros: A tutorial for
  social work and youth prevention science researchers.
\newblock {\em Journal of the Society for Social Work and Research},
  9(4):721--740, 2018.

\bibitem{modelingzeromodifiedcount}
Brian Neelon, A.~James O'Malley, and Valerie~A. Smith.
\newblock Modeling zero-modified count and semicontinuous data in health
  services research part 1: background and overview.
\newblock {\em Statistics in Medicine}, 35(27):5070--5093, 2016.

\bibitem{liu2019statistical}
Lei Liu, Ya-Chen~Tina Shih, Robert~L Strawderman, Daowen Zhang, Bankole~A
  Johnson, and Haitao Chai.
\newblock Statistical analysis of zero-inflated nonnegative continuous data.
\newblock {\em Statistical Science}, 34(2):253--279, 2019.

\bibitem{bouchaudprice}
Jean-Philippe Bouchaud.
\newblock {\em Price impact. Encyclopedia of Quantitative Finance. 2010}.
\newblock Wiley. AQ4.

\bibitem{webster2023handbook}
Kevin~T Webster.
\newblock {\em Handbook of Price Impact Modeling}.
\newblock Chapman and Hall/CRC, 2023.

\bibitem{patzelt2018universal}
Felix Patzelt and Jean-Philippe Bouchaud.
\newblock Universal scaling and nonlinearity of aggregate price impact in
  financial markets.
\newblock {\em Physical Review E}, 97(1):012304, 2018.

\bibitem{toth2015equity}
Bence Toth, Imon Palit, Fabrizio Lillo, and J~Doyne Farmer.
\newblock Why is equity order flow so persistent?
\newblock {\em Journal of Economic Dynamics and Control}, 51:218--239, 2015.

\bibitem{bucci2018slow}
Fr{\'e}d{\'e}ric Bucci, Michael Benzaquen, Fabrizio Lillo, and Jean-Philippe
  Bouchaud.
\newblock Slow decay of impact in equity markets: insights from the ancerno
  database.
\newblock {\em Market Microstructure and Liquidity}, 4(03n04):1950006, 2018.

\bibitem{donier2015fully}
Jonathan Donier, Julius Bonart, Iacopo Mastromatteo, and J-P Bouchaud.
\newblock A fully consistent, minimal model for non-linear market impact.
\newblock {\em Quantitative finance}, 15(7):1109--1121, 2015.

\end{thebibliography}
\newpage
\appendix

\renewcommand{\thesection}{Appendix}

\section{Aggregated Impact}\label{appendix_impact}
Inspired by the work of F. Patzelt and one of us (JPB) \cite{patzelt2018universal}, we quantify the relationship between the aggregated execution imbalance and its impact on the price for a bin size $N$. In a similar way, let us define  the aggregate-imbalance impact for $N$ consecutive observed price changes:
\begin{equation}
    \mathcal{R}_N (\mathcal{I}_N) = \Biggl \langle   m_{t+i} - m_t | \mathcal{I}_N =\sum_{i=0}^{N-1} V_i^{ex,a} - V_i^{ex,b} \biggr \rangle.
\end{equation}

\begin{figure}[h!]
  \begin{subfigure}{.5\textwidth}
  \centering
    \includegraphics[width=\linewidth]{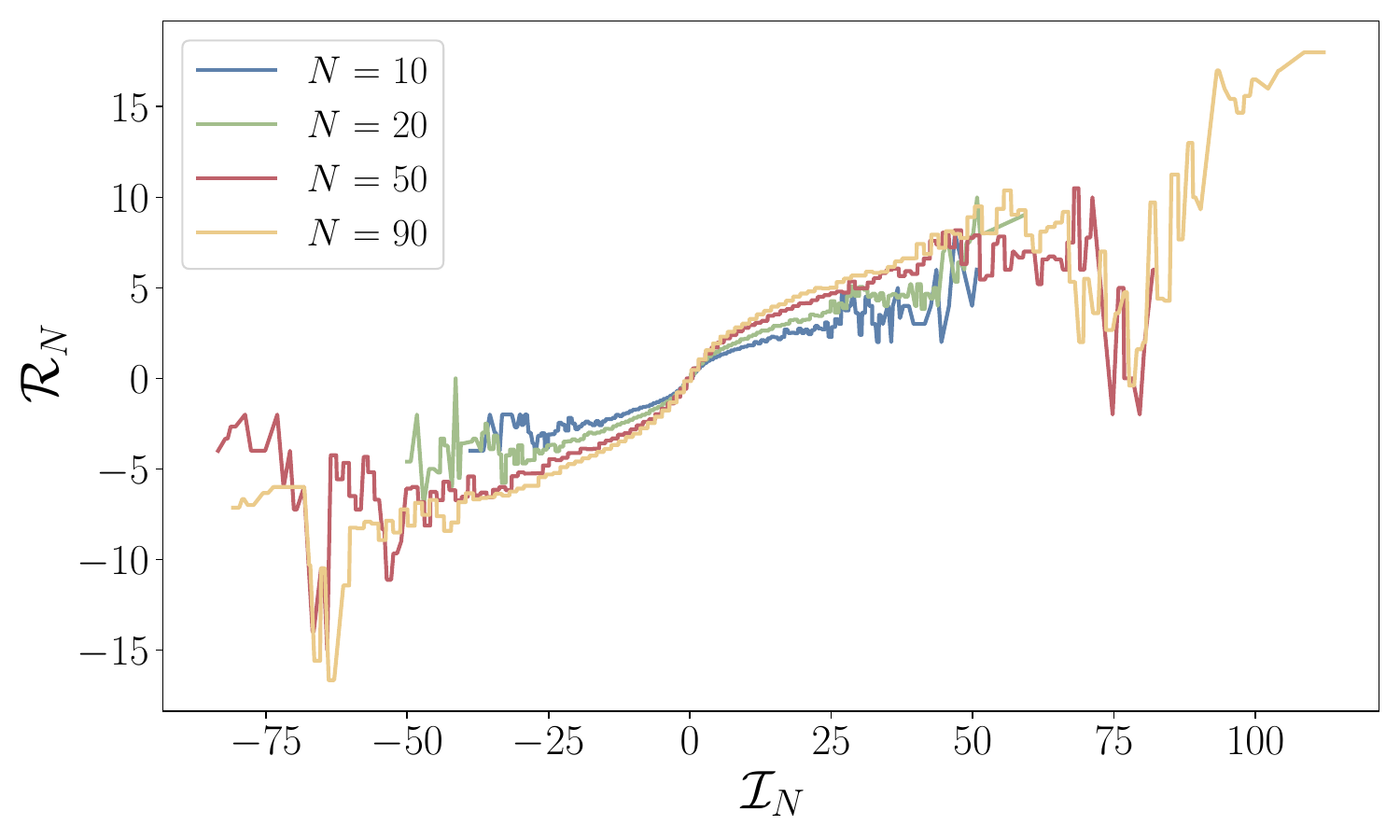}

  \end{subfigure}%
  \begin{subfigure}{.5\textwidth}
  \centering
    \includegraphics[width=\linewidth]{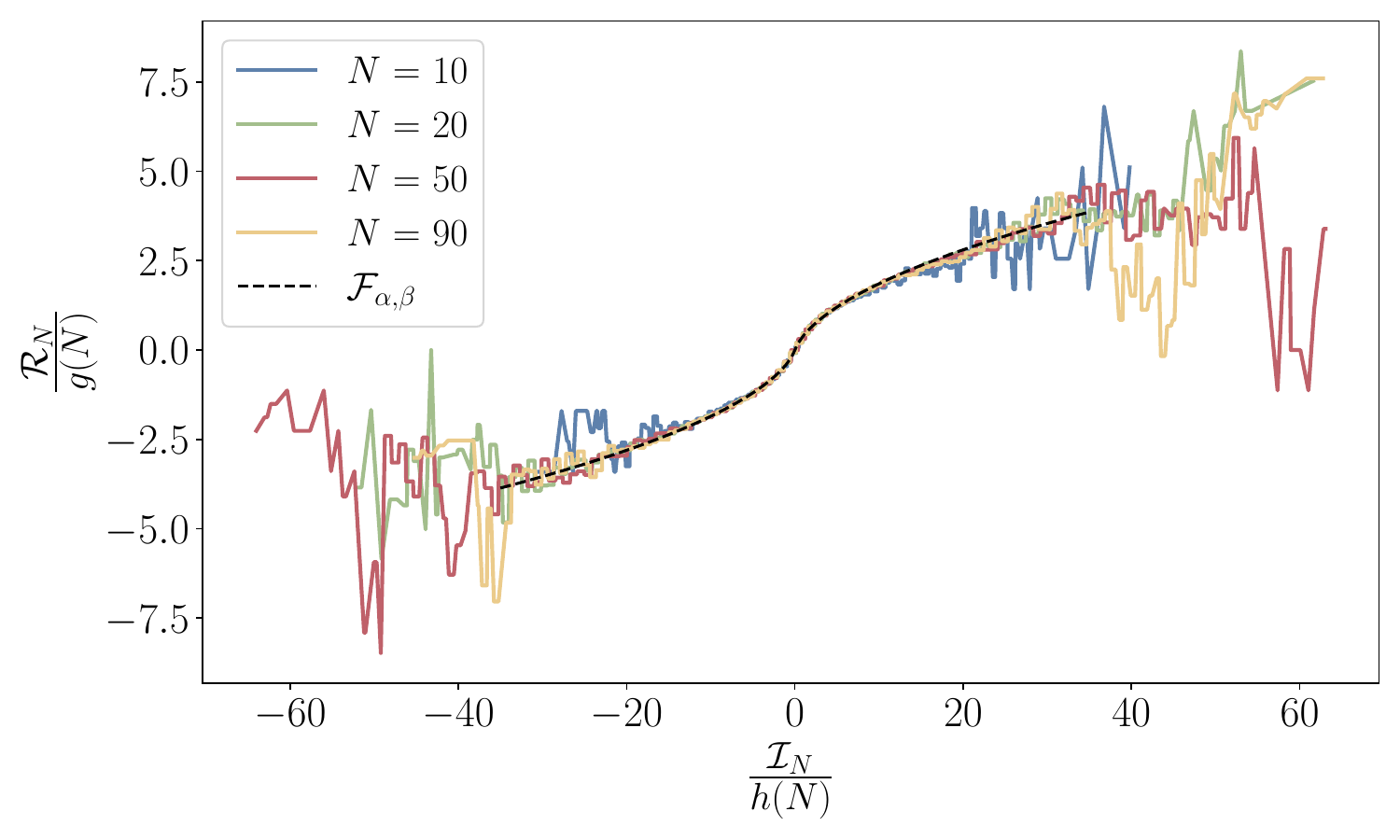}
  \end{subfigure}
  \caption{\textbf{Left}: Aggregated imbalance impact on raw data, for different bin sizes $N$ before the rescaling. \textbf{Right} : Aggregated imbalance impact on raw data, for different bin sizes $N$ after the rescaling. }
  \label{fig:scaling}
\end{figure}

As in \cite{patzelt2018universal}, we write:
\begin{equation}\label{scaling}
    \mathcal{R}_N (\mathcal{I}) = g(N) \mathcal{F_{\alpha, \beta}}\left(\frac{\mathcal{I}}{h(N)}\right),
\end{equation}
where $g(N)$ and $h(N)$ are the appropriate scaling of the return and the imbalance for a bin size $N$, and $\mathcal{F}_{\alpha, \beta}$ is a sigmoidal parametric function
\begin{equation}
    \mathcal{F}_{\alpha, \beta}(x) = \frac{x}{\left(1 + \left| x \right| ^\alpha\right)^{\frac{\alpha}{\beta}}}.
\end{equation}
\begin{figure}[htbp] 
\centering
  \centering
  \includegraphics[width=0.8\linewidth]{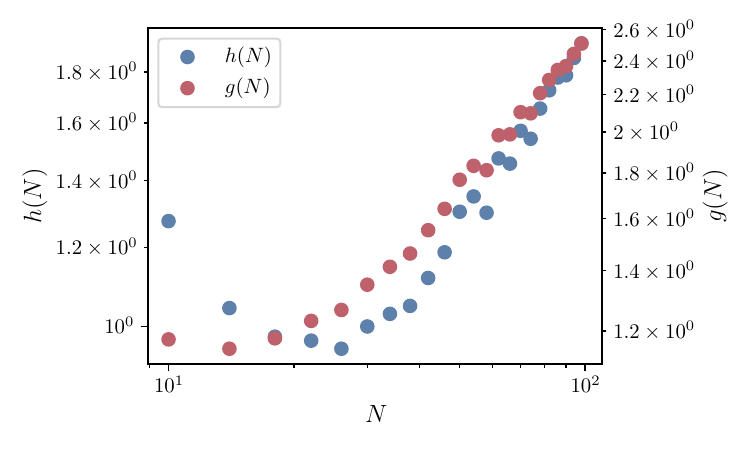}
\caption{Evolution of the scaling of of the impact and of the return. Starting from $N=20$, the evolution of the scaling is stable.}
\label{fig:scales_h_g}
\end{figure}
After calibrating of the parameters of \eqref{scaling}, the rescaled aggregated impact is the same for all the bin sizes $N$, as one can see in Figs. \ref{fig:scales_h_g} and \ref{fig:scaling}.

It is interesting to note that the universality of the aggregate impact holds even for price change-by-price change data, although the scaling of the returns and the impact no longer follows a pure power law.

\end{document}